# Spectral responsivity and photoconductive gain in thin film black phosphorus photodetectors

*Junjia Wang, Adrien Rousseau, Elad Eizner, Anne-Laurence Phaneuf-L'Heureux, Léonard Schue, Sébastien Francoeur, and Stéphane Kena-Cohen*

Department of Engineering Physics, Polytechnique Montréal, Montréal, Québec, H3C 3A7, Canada

ABSTRACT: We have fabricated black phosphorus photodetectors and characterized their full spectral responsivity. These devices, which are effectively in the bulk thin film limit, show broadband responsivity ranging from <400 nm to the ~3.8 μm bandgap. In the visible, an intrinsic responsivity >7 A/W can be obtained due to internal gain mechanisms. By examining the full spectral response, we identify a sharp contrast between the visible and infrared behavior. In particular, the visible responsivity shows a large photoconductive gain and gate-voltge dependence, while the infrared responsivity is nearly independent of gate voltage and incident light intensity under most conditions. This is attributed to a contribution from the surface oxide. In addition, we find that the polarization anisotropy in responsivity along armchair and zigzag directions can be as large as $10^3$ and extends from the band edge to 500 nm. The devices were fabricated in an inert atmosphere and encapsulated by $Al_2O_3$ providing stable operation for more than 6 months.

Photodetectors are ubiquitous in a wide range of scientific and industrial applications spanning security, process control, sensing and spectroscopy, imaging, defense, and astronomy. Silicon and InGaAs are well-established semiconductors for visible and and short-wave infrared (SWIR) detection[1]. Although narrow gap semiconductors such as PbS, InAsSb, HgCdTe, and quantum-well infrared (IR) photodetectors offer good performance in the mid-wave and long-wave infrared as compared to thermal detectors, they are costly, difficult to directly integrate within various optoelectronic platforms and generally require low-temperature operation[2]. For these reasons, there continues to be a need for novel materials and device architectures allowing for ease of integration and efficient detection in the IR.

Recently, monolayers of two-dimensional (2D) layered materials such as graphene and transition metal dichalcogenides (TMDCs) have emerged as attractive materials for optoelectronic devices due to their excellent optical, electrical and mechanical properties and the ease with which they can be integrated within different material platforms[3-5]. Graphene has a vanishing bandgap that allows for spectrally wide absorption and consequently broadband photodetection[6]. Furthermore, graphene's high carrier mobility makes it suitable for ultrafast operation[7]. However the responsivity of graphene is limited due its weak optical absorption coefficient[8]. For this reason, direct bandgap 2D materials such as the TMDCs molybdenum disulfide and tungsten diselenide have also been explored for use in photodetectors. In a field-effect transistor (FET) geometry, these materials exhibit high on-off ratios and excellent current saturation characteristics[4, 9], but their bandgaps are only suitable for UV, VIS, and SWIR regions of the spectrum[10] where excellent alternatives already exist.

Another layered material, black phosphorus (BP) has been "rediscovered" in recent years as an anisotropic material for optoelectronic and electronic applications[11]. This semiconducting material, characterized by its puckered crystal structure, can show a hole mobility up to 1000 $cm^2V^{-1}s^{-1}$ and large FET on/off ratios at room temperature when used as a channel material[11-15]. The atomic crystal structure of black phosphorus consists of layered $sp^3$-hybridized phosphorus atoms with distinct configurations along two directions, *i.e.* armchair (AC) and zigzag (ZZ), which lead to strong in-plane anisotropic physical properties, such as thermal conduction[16, 17], carrier transport and optical absorption[18-22]. In addition, the bandgap of BP is highly tunable by varying the number of layers. Thick multilayer BP flakes have a direct band gap of ~0.3 eV[13, 23], which is well-suited for SWIR and mid-wave IR photodetectors[24-37]. This bulk limit occurs for thicknesses above ~20 layers. By reducing the number of layers, the band gap can be increased up to ~1 eV, corresponding to the monolayer limit. As a result of these extraordinary properties, broadband photodetectors based on BP are being extensively studied. For example, Guo *et al.* have demonstrated photodetection at 532 nm and 3.39 µm using BP FET transistors with responsivities of up to 82 A/W at low incident intensities. This high responsivity was attributed to internal gain due to the photogating effect[38]. Indeed, internal gain mechanisms have been shown to lead to very large responsivities of $10^5$ A/W and $10^6$ A/W at low power for BP photodetectors in the UV and visible, respectively[30, 39]. Waveguide-integrated detectors have also been demonstrated by Youngblood *et al.* with a responsivity of 657 mA/W in the telecommunication band at an operation frequency of 3 GHz operation[40]. One of BP's most unique properties is the large tunability of its bandgap with an applied vertical electric field. Recently, a tunable BP photodetector was demonstrated with responsivities of 518, 30 and 2.2 mA/W at 3.4, 5 and 7.7 µm, respectively at 77 K[41].

Although BP photodetector research is now well-established, previous work on BP photodetectors has mostly focused on characterization in the visible/SWIR (<1700 nm) or single-wavelength responsivities. Here, we demonstrate thin film broadband BP photodetectors based on a FET structure and characterize the responsivity and photoconductive gain from 400 nm to 3.8 µm in order to paint a full picture of the detector response and identify the origin of internal gain mechanisms. We find distinct responsivity behavior in the visible and NIR. In the visible region of the spectrum, the responsivity is strongly doping and power-dependent. We find that at low incident intensity (~0.5 mW/cm$^2$ at 580 nm), internal gain mechanisms lead to responsivities as high as ~ 7 A/W. At higher powers corresponding to the linear operation regime, the responsivity drops to < 0.1 A/W. In addition, we find that at low power, the polarization anisotropy of the responsivity disappears for λ < 800 nm. It turns out, however, that the vanishing polarization anisotropy is a consequence of the power-dependent photoconductive gain. The polarization anisotropy is significantly increased in the IR, reaching ~10$^3$.

The photodiode architecture is schematically shown in Figure 1a. The devices were fabricated on a p-doped Si substrate with 90 nm SiO$_2$ oxide layer. First, BP was exfoliated onto the substrate using PVC tape (SPV 224PR-M, Nitto Denko)[42]. The thickness of the flakes (30 nm and 60 nm were measured by atomic force microscopy (AFM, Bruker Dimension FastScan). All exfoliation and transfer processes were performed in an nitrogen-filled glovebox to protect BP from oxidation. Interdigitated Cr/Au (10/60 nm) electrodes were deposited by vacuum thermal evaporation to form source and drain contacts, with a channel length of 2 µm. Finally, 1 nm of thin aluminium was evaporated to form a natural Al$_2$O$_3$ top oxide layer and protect BP. An optical microscope image of the completed 30 nm-thick device is shown in Fig. 1b. The active area is ~320 um$^2$. Polarization-resolved Raman spectroscopy was used to determine the axis of the flakes (Fig. 1c, see Methods for details). The interdigitated electrodes were oriented along the ZZ direction of the BP flake to ensure current flow along the AC direction during operation. Given that BP is unstable in air in particular with the presence of both oxygen and water[43], an effective passivation technique is essential. Here, the devices are encapsulated with a thin layer of $Al_2O_3$ which reduces degradation of the BP active layer[44].

Figure 1d shows the transfer characteristic of the 30 nm-thick device measured under ambient conditions at a source-drain voltage, $V_{DS}$ = 1 mV. The device is p-type with an intrinsic resistance ~300 Ω. We note that the on/off ratio for the range of gate voltages shown is only ~15, which is typical of thicker BP FETs due to the relatively high conductance of the naturally doped bulk crystal and the small out-of-plane screening length[28, 45, 46].

To measure the responsivity across the entire spectral range, two separate light sources are used. For 400–1700 nm, a supercontinuum (SC) laser (WhiteLase SC450-4, Fianium) is focused onto the input slit of a monochromator to provide tunable single wavelength excitation of the sample with an irradiance of 10-2000 mW/cm$^2$ and a ~10 nm bandwidth. For IR (1000–4000 nm) characterization, a Fourier-transform infrared spectrometer (FTIR) in step-scan mode with a globar light source is used. Applying an additional LED light during the FTIR measurement confirmed that the shape of the responsivity was independent of the incident light intensity. This verification is important in cases where intensity-dependent photoconductive gain can lead to artifacts in the FTIR response. Absolute responsivities are obtained by comparing the measured photocurrent to that obtained using NIST-traceable calibrated reference detectors spanning 400–1700 nm. Relative responsivity is obtained for the FTIR measurement by comparing the photocurrent to that from a DLaTGS IR detector possessing a flat responsivity over the measured spectral range. The absence of gain under FTIR measurement conditions simplifies the analysis and allows us to stitch the absolute responsivity measured with the SC source to the relative responsivity measured with the FTIR. For most measurements, the stitching wavelength is chosen to be λ = 1500 nm, which is the wavelength where gain is also minimal for the SC measurement. Under all measurement conditions, an ultra-broadband wire-grid polarizer was used to align the polarization along the AC or ZZ directions and all-reflective optics were used to avoid chromatic aberration.

Figure 2a shows the measured responsivity as a function of the light intensity at a source-drain voltage $V_{DS}$ = 50 mV. The gate voltage was fixed at $V_{GS}$ = -20 V, which corresponds to the FET "on" state (Fig. 1d). The irradiance of the supercontinnum source was varied between 15 mW/cm$^2$ to 1 W/cm$^2$ (reported at λ = 900 nm; see inset for the relative spectral density). From the results, we immediately notice a clear spectral dependence of the responsivity in the VIS and NIR range. The responsivity reaches a maximum of 6 A/W at λ=580 nm, which corresponds to the wavelength for which the illumination source has the lowest irradiance. In contrast, the NIR responsivity (1000-1500 nm) remains almost constant around 2 x 10$^{-3}$ A/W. At the highest measurement power (>400 mW/cm$^2$), photoconductive gain, which is at the origin of the change in responsivity with illumination intensity, is greatly reduced and the response becomes nearly independent of the pump power. Fig. 2b shows the responsivity of the 60 nm-thick sample at 2W/cm$^2$ under conditions where the minimum in the illumination power at λ=580 nm is eliminated by using a different monochromator grating and the device is additionally flooded with a white light LED bias to completely eliminate photoconductive gain. This responsivity corresponds to the intrinsic, linear responsivity of the BP photodetector.

In p-doped BP, it is expected that electron traps due to the presence of defects and surface states are at the origin of phoconductive gain. The gain is given by $G = \tau_t/T$, where $\tau_t$ is the trap lifetime and $T$ is the carrier transit time across the device. Note that the former depends strongly on the fabrication conditions, while the latter depends on both the measurement conditions ($V_{DS}$) and device structure (channel length). This strong variation explains the broad range of *G* measured across previous reports. For high illumination intensities, trap states that are at the origin of the photoconductive gain are filled and the responsivity closely follows the previously measured absorption of BP[47]. As expected, the responsivity drops around 3.7 μm which corresponds to the bandgap of BP. We find that there is a gradual decrease of the responsivity when moving from 450 nm to the bandgap. This variation is most important in the visible and decreases reduced beyond λ~1.5 μm. We also observe dips in the IR responsivity corresponding to atmospheric water and $CO_2$ absorption, which were not explicitly removed.

To evidence the role of photoconductive gain and its spectral dependence, we explicitly measured the responsivity of the 30 nm-thick detector as a function of input irradiance in Fig. 2c. The measurements are performed at three different wavelengths: 570 nm, 1320 nm and 1700 nm. Here, the power is reported at the corresponding wavelengths. For visible λ = 570 nm illumination, the responsivity decreases from 5 A/W to 0.13 A/W as the irradiance increases from ~1 mW/cm$^2$ to 300 mW/cm$^2$. In contrast, the responsivities remain almost constant for 1320 nm and 1700 nm illumination. This confirms that the large photoconductive gain in the visible is not solely due to the minimum incident power spectrum (inset of Fig. 2a). We suspect that photogating from various forms of surface oxide may be responsible for this behavior. Although the most stable phosphorus oxide, $P_2O_5$, is transparent in the visible, various suboxides $P_4O_n$, with n<8 have bandgaps in the ~2 eV range[48]. Raman measurements of the full device with $Al_2O_3$ surface layer (see Supplementary Information), show an $A_g^1/A_g^2$ ratio of ~0.3, which is also indicative of some oxidation. Similarly, gain in $In_2Se_3$ photodetectors has previously been attributed to photogating from the natural surface oxide[49]. In the NIR (e.g. 1320 nm and 1700 nm), photoconductive gain only becomes apparent at irradiances <0.1 mW/cm$^2$.

To confirm that photothermal effects play a neglibile role under our measurement conditions, we examine the I-$V_{DS}$ characteristic in the dark and under illumination. Fig. 3a and b show the difference between the current under illumination at λ = 700 nm and λ = 1550 nm for irradiances ranging from 0.6 W/cm$^2$ to 3 W/cm$^2$, $I_{PH}$, and the dark current, $I_D$. At both wavelengths, the photocurrent is along the direction of the applied bias, regardless of the input power, and increases with increasing electric field. This behavior is consistent with the photoconductive mechanism, in which illumination leads to an increase in the channel conductivity.

To examine the effect of the gate voltage on the responsivity, we measure the responsivity as a function of gate voltage. Figure 4a shows the measured responsivity of the 30 nm-thick device as a function of the gate voltage at a source-drain voltage $V_{DS}$ = 50 mV and an irradiance of 1W/cm$^2$ measured at λ = 900 nm. The responsivity reaches its maximum at $V_{GS}$ = -20 V, and we find a maximum responsivity of 0.06 A/W at λ = 580 nm, where the incident power is the lowest and a low responsivity in the 1.1 – 1.4 µm range with a minimum of responsivity of 2 x 10$^{-3}$ A/W. These results are consistent with those observed in Fig. 2a. In the IR, the responsivity is ~ 2 x 10$^{-3}$ A/W around 2-3 µm. In particular, the gate voltage only affects the visible part of the responsivity (λ < 1000 nm). Further measurements on the 60 nm thick-device (see Supplementary Information) show an initial rise in photocurrent near -10V, followed by a drop in photocurrent near -35V as $V_{GS}$ is made increasingly more negative. This gate dependence is anticipated due to the interplay between the Fermi level shift and photoconductive gain. When the gate voltage is reduced, the Fermi level approaches the valence band, which leaves most electron traps empty and allows for high gain. This is the phenomenon observed in Fig. 4. For very large p-dopings (i.e. $V_G$ < -35V), however, large Schottky barriers can form at the contacts, which increases recombination near the contacts and reduces the photocurrent (see Supplementary Information).

Figure 4b shows the polarization anisotropy in the detector response. In this case, the responsivity was separately measured for polarizations along the AC and ZZ directions at a high irradiance (1W/cm² at 900 nm). The AC direction is characterized by a larger field-effect mobility and a higher absorption coefficient in the IR. Indeed, we find that in the IR, the responsivity along the AC direction is ~1000 larger than that along the ZZ direction. The anisotropy is reduced at lower illumination powers or when approaching the visible part of the spectrum. This is due to an interplay between photoconductive gain and anisotropy in the absorption coefficient. When the optical absorption is low, the photoconductive gain is higher, which artificially decreases the influence of the anisotropy on the responsivity. This effect is particularly important in the visible where we find the strongest influence of gain. The discrepancies found amongst previous reports in both the value of the anisotropy and the spectral range where it occurs can be largely explained by the sensitivity of the responsivity on the irradiance.

To measure the noise equivalent power of the detector, noise spectrum was first measured with a spectrum analyzer and normalized to 1 Hz. Note that the detector dark current and preamplifier noise were taken into consideration. The measured power spectral noise is converted to the noise current spectral density by the equation $NEP = P_{noise}/\sqrt{BW}$, where $P_{noise}$ is the integrated input noise power and BW is the bandwidth taken into consideration. Detailed measurement steps can be found in reference [50]. The NEP is measured to be ~12 pW Hz$^{-1/2}$ at 580 nm and ~470 pW Hz$^{-1/2}$ at 2.5 μm ($V_{DS}$ = 50 mV, $V_{GS}$ = -20V). The thicker BP layer, as compared to previous reports, leads to a slightly higher NEP.[41]

In summary, we have studied the full spectral response of thin film BP photodetectors with a detection window extending from 400 nm to ~3.8 μm. We find distinct behaviors for the responsivity in the visible and the IR, which we tentatively attribute to photogating effects originating from the surface oxide layer. In the visible, the devices reach ~7 A/W responsivity at -20 V back gate voltage. We have shown that the responsivity of the anisotropy can reach up to $10^3$ in the IR and extend much further in the visible than previously reported. This very large polarization anisotropy is relevant for a number of applications, including integrated polarimetry. The encapsulated devices have shown stable operation for more than 6 months.

METHODS

**Device fabrication**

The fabrication process is the following. We start with a commercial p-type silicon substrate (Silicon/Silicon dioxide (90 nm) wafers from Graphene Supermarket) with a 90 nm oxide layer on top. The substrate was first dehydrated and coated with hexamethyldisilizane (HMDS) to passivate the surface (HMDS Prime Oven, Yield Engineering System Inc.). Next, BP (Black Phosphorus, HQ Graphene) is mechanical exfoliated onto the substrate using a nitto tape inside a glovebox followed by high temperature annealing to reduce tape residues. The BP crystalline orientation was identified by polarization-resolved Raman spectroscopy. Then the electrodes patterns are defined by electron beam lithography (EBL, Raith eLine) using PMMA resist. 10 nm of chromium and 60 nm of gold are thermally evaporated (EvoVac, Angstrom Engineering). Finally, after the lift off, a 1 nm of aluminium is deposited on top by thermal evaporator.

**Polarization-resolved Raman spectroscopy**

We performed a polarization-resolved Raman spectroscopy measurement in order to identify the orientation of the crystal axes of the flake. The data was corrected[51] for absorption anisotropy, birefringence and dichroism between zigzag and armchair directions[52] (enhancement of 1.22 of the ZZ contribution compared to AC for a 30 nm thick flake) and interference in the 90 nm thick

SiO$_2$ layer. It was then fitted with a model considering the Raman tensor elements of both axes, their relative phase and an angle offset between the laboratory and crystal axes[53].

**Device characterization**

The experimental setup is shown in Supplementary Section Fig. S1. Finished devices were mounted onto a translation stage and contacted by electrical probes. Current–voltage measurements were taken in a two-probe configuration using a source meter (2614B, Keithley). For the visible and near IR part, the spectral response was characterized by focusing light beam from a supercontinuum laser. A monochromator (SP-500i, Princeton Instruments) with long pass filters to avoid higher-order diffraction harmonics is used to provide a tunable single wavelength excitation and a chopper is placed after to modulate the light at a frequency of 500 Hz. The supercontinuum laser repetition rate is 40 MHz, corresponds to quasi-CW excitation under our measurement conditions. For the IR part of the measurement, the FTIR's internal Globar light source was used. The resulting photocurrent was then sent to a preamplifier (SRS570, Stanford Research Systems) followed by a lock-in amplifier (MFLI, Zurich Instruments) and subsequently returned to the external detector interface of the FTIR. For polarization-resolved measurements, a wire grid polarizer was placed before the reflective objective lens. The built-in DLaTGS pyroelectric detector was used to measure the relative spectral intensity of the source.

AUTHOR INFORMATION

**Corresponding Author**

*s.kena-cohen@polymtl.ca

**Notes**

The authors declare no competing financial interest.

ACKNOWLEDGMENT

The authors would like to acknowledge helpful discussions with Mahmoud Atalla. S.K.C. acknowledges support from the Canada Research Chairs program. S.K.C. and S.F. gratefully acknowledge funding for this work from NSERC Strategic Grant STPGP-506808.

**FIGURES**

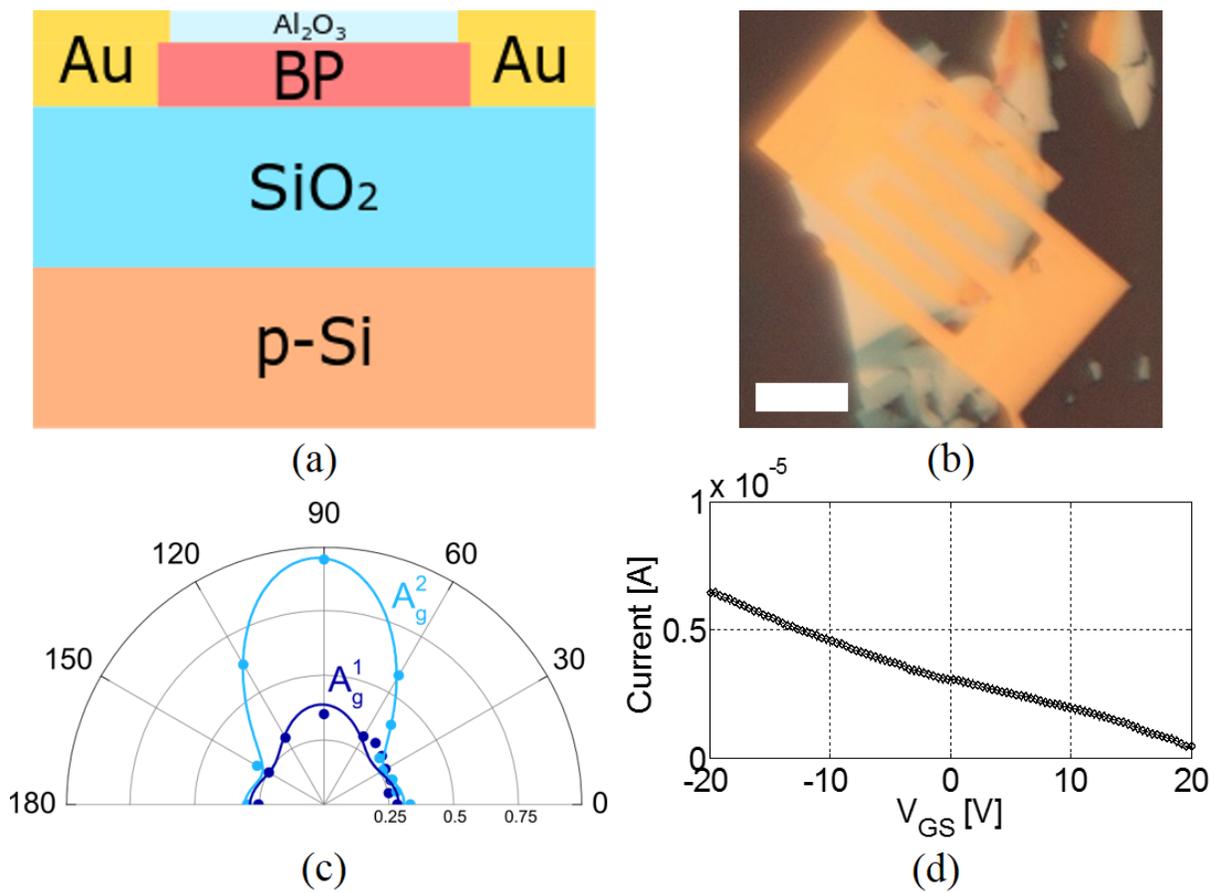

Figure 1. (a) Schematic BP PD. P-Si: p-doped silicon. (b) Optical image of the BP PD. Scale bar: 10 μm. (c) Polarization-resoloved Raman plots of $A_g^1$ and $A_g^2$ peaks. (d) Source-drain current as a function of gate voltage.

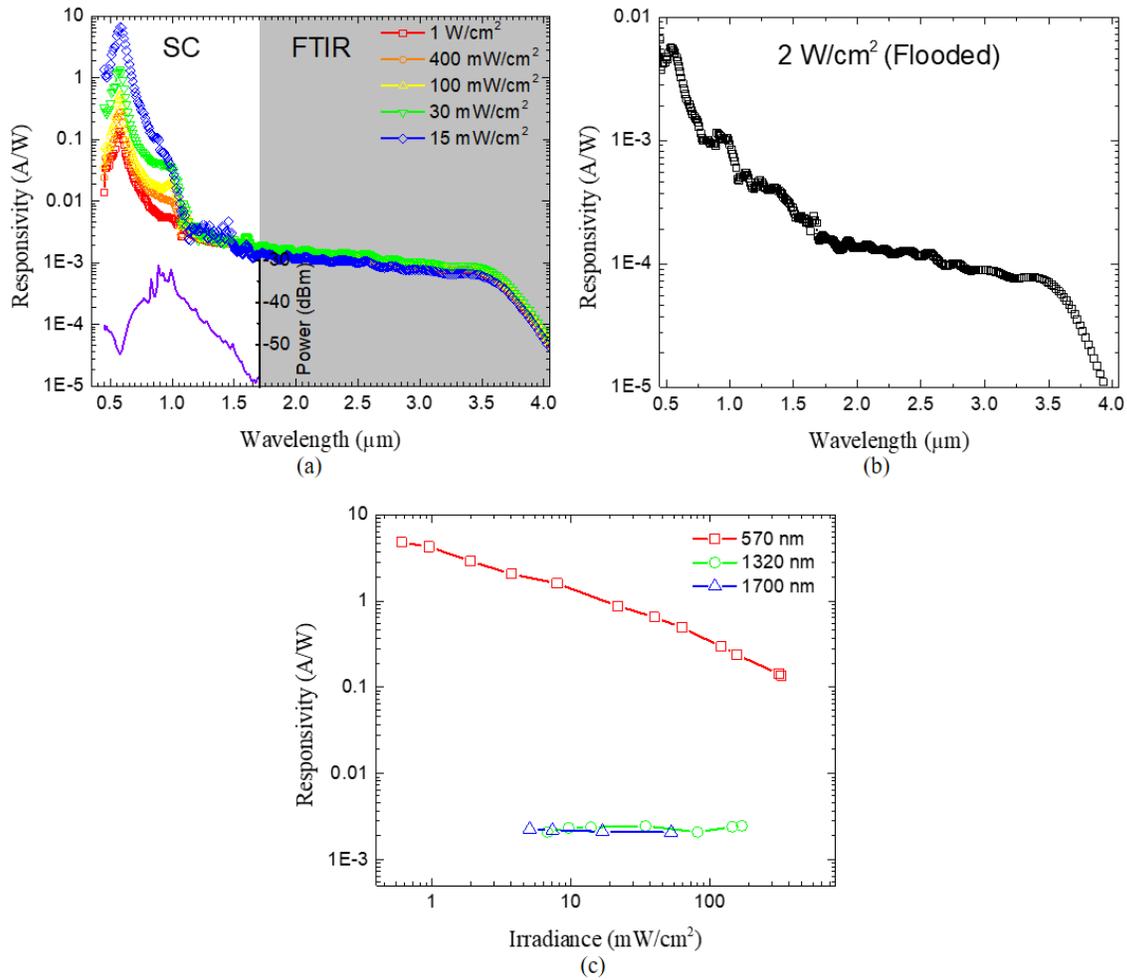

Figure 2. (a) Responsivity of the 30 nm-thick photodetector as a function of irradiance. The inset shows the illumination power as a function of wavelength. The shaded region delineates the monochromator and FTIR-based characterization techniques. The reported irradiances are at 900 nm. (b) Responsivity of the 60 nm-thick detector, with a grating eliminating the visible dip in power and additionally flooded with a white LED. This responsivity corresponds to the intrinsic, linear response of the BP photodetector. (c) Single-wavelength responsivity of the device as a function of irradiance at the indicated wavelengths. In all panels, the measurements were performed at -20 V gate voltage.

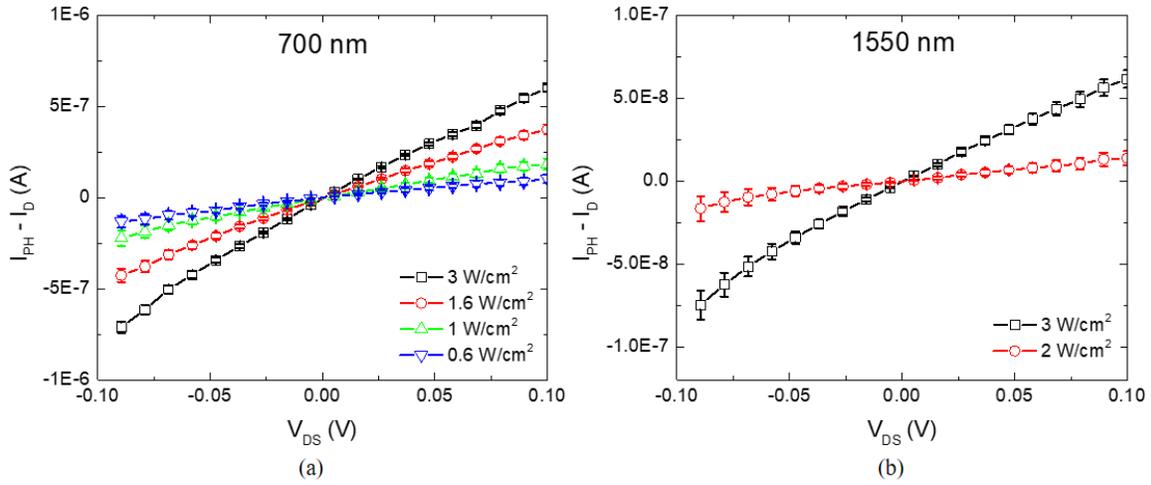

Figure 3. Photocurrent, which corresponds to the difference in current under illumination, $I_{PH}$, and dark, $I_D$, as a function of the drain-source voltage, $V_{DS}$, of the 60 nm-thick device at (a) 700 nm (b) 1550 nm. The measurements were performed at various irradiance.

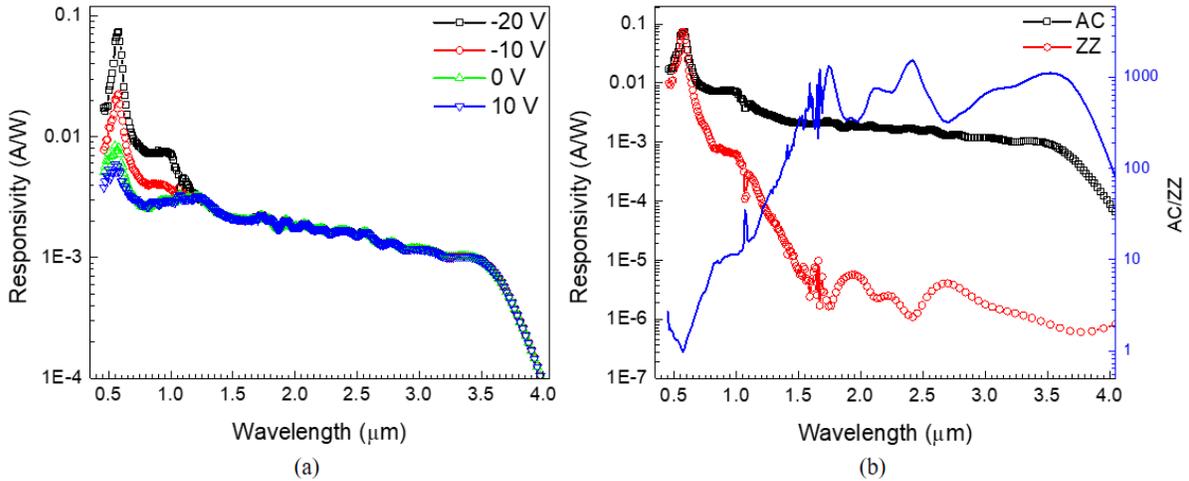

Figure 4. Spectral dependence of the responsivity as a function of (a) gate voltage and (b) polarization measured with an irradiance of 1 W/cm$^2$ at 900 nm. The blue line (right axis) gives polarization anisotropy of the responsivity.